\documentclass[australian,english]{article}
\usepackage[T1]{fontenc}
\usepackage[latin9]{inputenc}
\usepackage{mathrsfs}
\usepackage{amsmath}
\usepackage{amssymb}
\usepackage{esint}

\makeatletter

\newcommand{\noun}[1]{\textsc{#1}}

\makeatother

\usepackage{babel}
\begin{document}
\title{Completing the quantum ontology with the electromagnetic zero-point
field}
\date{}
\author{Luis de la Peña and Ana María Cetto}

\maketitle
Instituto de Física, Universidad Nacional Autónoma de México, Mexico
City, Mexico
\begin{abstract}
This text begins with a series of critical considerations on the initial
interpretation of quantum phenomena observed in atomic systems. The
bewildering explanations advanced during the construction of quantum
mechanics are shown to have distanced the new theory from the rest
of scientific knowledge, by introducing indeterminism, acausality,
nonlocality, and even subjectivism as part of its interpretative framework.
The conclusion drawn from this unsatisfactory interpretative landscape
is that quantum mechanics lacks a key ontological ingredient. Arguments
are given in favour of the random zero-point radiation field (\noun{zpf})
as the element needed to complete the quantum ontology. The (wave-mediated)
quantum stochastic process is shown to be essentially different from
Brownian motion, and more amenable to an analogy with the hydrodynamic
case. The new perspective provided by the introduction of the \noun{zpf}
is used to explain some salient features of quantum systems, such
as the stationary atomic states and the transitions between them,
and the apparent nonlocality expressed in the entangled states. Notably,
the permanent presence of the field drastically affects the dynamics
of the (otherwise classical) particle, which eventually falls under
the control of the field. This qualitative change is reflected in
the transition from the initial classical description in space-time,
to the final quantum one in the Hilbert space. The clarification of
the mechanism of quantization leads us to consider the possibility
that a similar phenomenon occurs in other physical systems of corpuscles
subjected to an oscillating background, of which the walking-droplet
system is a paradigmatic example.
\end{abstract}

\section{Introduction}

A careful study of quantum mechanics (\noun{qm}) as it is systematically
presented and understood in present-day publications, from the dominant
Copenhagen perspective or any of the alternatives---which can be
counted by the dozen---leads to the conclusion that it must be indeed
an incomplete theory. Only an \foreignlanguage{australian}{incomplete}
theory can be 'completed' with such a plentiful variety of alternatives,
some even in clear contradiction amongst them. But rather than the
incompleteness considered by Einstein ---any statistical description
is incomplete by nature---we refer here to an essential ingredient
that is missing. This assertion, quite significant for a theory that
is considered to be the basis of half (if not more) of modern physics,
gets reaffirmed by a critical analysis of the origin of a few of its
most representative and fundamental postulates---some implicit, others
explicit.

One such postulate that calls for immediate attention is related to
the early discovery by Heisenberg of the hazardous motions of the
electrons. Being unaware of any known explanation, Heisenberg proposed
to consider such motions as a trait of \textit{\emph{nature}} and
\textit{\emph{postulated}}\emph{ }that the electron (as any other
quantum particle) has an inherently indeterministic and acausal behaviour.
This postulate was blindly accepted by the physics community of the
time, and was never subjected to experimental corroboration, not even
after the advent of quantum field theory, which inherited it. It is
interesting to note that even the community of philosophers of science
involved in quantum theory, otherwise so demanding, accepted Heisenberg's
postulate as a significant part of reality. With this step, the phenomenological
description---of the \textit{save-the-phenomenon} type---provided
by quantum mechanics was adopted and taken for a fundamental, first-principles
construction.

On the other hand, the Schrödinger description and variants thereof
contain, either tacitly or explicitly, a nonlocal element that is
present in all instances except for the (rather unphysical) case of
constant probability density in the entire space. Physicists became
accustomed to use (even inadevertently) quantum nonlocality and accept
nonlocal behaviour as a quantum trait; of recently this point has
taken on increased importance. Being locality a property that is expected
of any fundamental physical theory, the conceptual---and philosophical---difficulties
with usual quantum theory continued to accumulate.

On top of this, we recall the description of the atom à la Heisenberg
as an entity that lives in an abstract, mathematical space---a Hilbert
space, a well defined mathematical structure---with no indication
at all of what is taking place in real space-time. Such a description
leaves us without a meaningful and transparent picture of what the
atom is and what its electrons are doing. A further difficulty is
related to the introduction of 'quantum jumps' between states, which
were (and still are, to a large extent) taken as a capricious quantum
trait, not amenable to further analysis. Strictly speaking, we rely
on a powerful formal description of the atom, with no associated intelligible
picture of it.

Despite the fact that the difficulties mentioned here (among others)
strongly suggested the need to adopt a statistical perspective of
the quantum phenomenon \cite{Bal98}, this possibility was dismissed
in general (and adamantly opposed by the Copenhagen school in particular),
opening the door to another infelicitous ingredient, the observer.
The introduction of this \textit{active} character in order to 'explain'
the reduction of the characteristic quantum mixtures to the pure states
observed, added a subjective ingredient to the already odd quantum
scheme.

\section{The missing ingredient}

From a narrative as the one just presented, one concludes indeed that,
for historical reasons, in the process of construction of quantum
mechanics some fundamental ingredient was left aside. The absence
of an appropriate ontological element turned the physical situation
into a mystery. The missing ingredient being so fundamental, its disclosure
should, as a minimum, explain the quantum motions that go under the
indeterministic name of quantum fluctuations, as well as the origin
of the apparent nonlocality, the existence of stationary states and
the spontaneous transitions between them.

Since all atomic constituents, whether charged or not, have an electromagnetic
structure, and electromagnetic interactions are ubiquitous in the
atomic world, a natural candidate to fill the ontological gap is the
random zero-point electromagnetic field (\emph{\noun{zpf}}). This
stochastic field was well known since its introduction into the quantum
world by Max Planck in 1912 \cite{Pl1911}, yet it was largely ignored
for decades. Taken as a real electromagnetic field that pervades the
entire space, one may envisage that any quantum particle with electromagnetic
properties is in permanent contact with it and therefore acquires
an essentially stochastic motion \cite{TWM63}. In addition, by serving
as a bridge that connects the individual particles of a system, this
field is expected to induce correlations between their motions even
when they do not interact directly, thus leading to an apparently
nonlocal behaviour, as is manifested e. g. in quantum entanglement.

More broadly, consideration of the \noun{zpf} as a fundamental ontological
constituent offers a range of possibilities to explain quantum phenomena.
An important one is the fact that the electron extracts energy from
the \noun{zpf}, which can make up for the loss of energy due to the
radiation of the accelerated charge. As noted by Nernst as early as
1916 \cite{Nernst1916}, this may explain a most intriguing and persistent
mystery, namely atomic stability: how is it that the electron is permanently
radiating, but the atom is stable? What's more: a precise compensation
of the mean energy radiated to the field by the mean energy absorbed
from it can take place only for certain very specific orbital motions,
offering in principle an \textit{\emph{explanation}}\emph{ }of atomic
quantization. Simple calculations or estimates taking into account
the known properties of the \noun{zpf} give support to these conjectures.
The theory based on the assumption of the \noun{zpf} as an essential
ingredient, called stochastic electrodynamics (\noun{sed}), is not
yet fully developed, but it has already produced a variety of positive
and promising results (\cite{Boyer}-\cite{TEQ} and references therein). 

One important lesson of the theory developed so far is that, as a
result of the permanent action of the background field on an otherwise
classical particle, a \textit{qualitative, irreversible change in
the dynamics} takes place: the particle ceases to behave classically
and acquires properties that are considered inherently quantum \cite{QSMF2021,FOOP2022}.
The effect of the field is not a mere perturbation, which means that
a perturbative approach to the \noun{sed} problem is doomed to produce
erroneous results in general. No perturbative calculation to any order
will give rise to a qualitatively different behaviour; the $new$
situation requires a \textit{new} description. And this is just the
quantum one. Briefly: \noun{qm} ceases to be a mechanical theory to
become an \emph{electrodynamic} theory.

The introduction by Planck of the zero-point field and its associated
energy $\mathcal{E}=\hbar\omega/2$ per mode, can be considered as
significant and groundbreaking as his introduction of the quantization
of the energy $\mathcal{E}=\hbar\omega$ interchanged between matter
and field, which led him to his famous blackbody formula. On one hand,
the zero-point term means a definitive departure from classical electromagnetism
by establishing a nonzero energy for the ground state of the field.
Further, it provides the basis for an understanding of the quantum
phenomenon, as proposed by \noun{sed}. The fact that the total energy,
integrated over the entire range of frequencies from 0 to $\infty$,
has an infinite value, has been used as an argument to deny the reality
of this field. This problem, however, is not unique to \noun{sed};
it is shared by quantum theory, which deals with it in the form of
vacuum fluctuations. The infinite energy of the vacuum is in fact
an open problem for cosmology, and different attempts to solve it
can be found in the literature (see e. g. \cite{San22}).

In this regard it is worth mentioning the proof by Unruh \cite{Unruh95}
that the black-hole evaporation process is insensitive to the high-frequency
regime. Unruh ascribes this to the time scales involved in the process,
which are inversely proportional to the mass of the black hole, hence
relatively long compared with Planck or atomic scales, and concludes
that 'if the state is the vacuum state at high frequencies, it remains
the vacuum.' In \noun{sed} (as in non-relativistic \noun{qm}) the
particles interact predominantly with modes of low frequencies, whence
the high-frequency modes have no effect on the dynamics; particles
are essentially transparent to them. Moreover, at frequencies higher
than (double) the Compton frequency, relativistic (high-energy) processes
such as particle creation and annihilation take place. Therefore in
the calculation of (nonrelativistic) radiative corrections, in which
the entire spectrum intervenes in principle, it is legitimate to introduce
a cutoff at the Compton frequency (as is usually done in nonrelativistic
\noun{qed}). 

Coming back to the conceptual, philosophical considerations, we conclude
that the mystery and magic that have accompanied the quantum world
over decades, may be dissolved in principle by considering the presence
of the \noun{zpf }as a real, physical field in permanent contact with
matter. Its introduction as an inseparable ingredient of the ontology
of any quantum system allows us to recover determinism, causality,
locality and objectivity. The corollary is that the classical and
the quantum worlds are not two distinct, separate worlds, each obeying
its own rules; there is a single world in which they coexist.

\section{On the nature of quantum stochasticity\label{SQM}}

A phenomenological theory called stochastic mechanics (alternatively,
stochastic quantum mechanics, \noun{sqm}) was initiated by the mathematician
Edward Nelson \cite{Nels66} with the purpose of describing quantum
mechanics as a stochastic phenomenon without the need to specify the
source of stochasticity. A more general formulation of \noun{sqm}
was developed later, which serves to describe the dynamics of two
distinct types of stochastic process, in the Markov approximation:
the classical, Brownian-motion type and the quantum one \cite{Pe69,Dice}. 

An important feature of \noun{sqm} is the appearance of two (statistical)
velocities on an equal footing: the flux (or flow) velocity $\boldsymbol{v}$
and the diffusive (or stochastic) velocity $\boldsymbol{u}$. These
basic kinematic elements for the description are obtained by averaging
over the ensemble of particles in the neighborhood of $\boldsymbol{x}$
at times close to $t$. When the time interval $\varDelta t$ is taken
small but different from zero, the two velocities are obtained, namely
(see e. g. \cite{Nels66,Dice})
\begin{equation}
\boldsymbol{v}(\boldsymbol{x},t)=\frac{\overline{\boldsymbol{x}(t+\Delta t)-\boldsymbol{x}(t-\Delta t)}}{2\Delta t}=\mathcal{\hat{D}}_{c}\boldsymbol{x},\label{3-2}
\end{equation}
with the systematic derivative operator given by
\begin{equation}
\mathcal{\hat{D}}_{c}=\frac{\partial}{\partial t}+\boldsymbol{v}\cdot\boldsymbol{\nabla},\label{3-4}
\end{equation}
and 
\begin{equation}
\boldsymbol{u}(\boldsymbol{x},t)=\frac{\overline{\boldsymbol{x}(t+\Delta t)+\boldsymbol{x}(t-\Delta t)-2\boldsymbol{x}(t)}}{2\Delta t}=\mathcal{\hat{D}}_{s}\boldsymbol{x},\label{3-6}
\end{equation}
with the stochastic derivative operator given by
\begin{equation}
\mathcal{\hat{D}}_{s}=\boldsymbol{u}\cdot\boldsymbol{\nabla}+D\boldsymbol{\nabla}^{2},\label{3-8}
\end{equation}
and the diffusion coefficient
\begin{equation}
D=\frac{\overline{(\Delta x)^{2}}}{2\Delta t},\label{3-10}
\end{equation}
assumed to be constant. The symbol $\overline{\left(\cdot\right)}$
denotes ensemble averaging.

The two time derivatives (\ref{3-4}) and (\ref{3-8}), applied to
the velocities (\ref{3-2}) and (\ref{3-6}), give rise to four different
accelerations, which are used to construct a couple of generic dynamical
equations. In the absence of an external electromagnetic field these
are the time-reversal invariant generalization of Newton's Second
Law, and the time-reversal non-invariant equation leading to the continuity
equation, respectively, 
\begin{eqnarray}
m\left(\mathcal{\hat{D}}_{c}\boldsymbol{v}-\lambda\mathcal{\hat{D}}_{s}\boldsymbol{u}\right)=\boldsymbol{f},\label{3-12}\\
m\left(\mathcal{\hat{D}}_{c}\boldsymbol{u}+\mathcal{\hat{D}}_{s}\boldsymbol{v}\right)=0,\label{3-14}
\end{eqnarray}
with $\lambda$ a free, real parameter, and $\boldsymbol{f=-\nabla V}$
the external force acting on the particle. Since the magnitude of
$\lambda$ can be absorbed into the value of $D$, one takes $\lambda=\pm1$.
The specific dynamical properties of the system strongly depend on
the sign of this parameter: $\lambda=-1$ implies an irreversible
dynamics, of the Brownian-motion type. By contrast, $\lambda=1$ implies
a reversible stochastic dynamics and leads after some algebra to the
Schrödinger-like equation 
\begin{equation}
-2mD^{2}\boldsymbol{\nabla}^{2}\psi(\boldsymbol{x},t)+V(\boldsymbol{x})\psi(\boldsymbol{x},t)=2imD\frac{\partial\psi(\boldsymbol{x},t)}{\partial t},\label{3-16}
\end{equation}
and its complex conjugate, where $\psi(\boldsymbol{x},t)$ is a complex
function whose squared modulus is the density distribution 
\begin{equation}
\rho(\boldsymbol{x},t)=|\psi(\boldsymbol{x},t)|^{2},\label{3-18}
\end{equation}
and 
\begin{equation}
\boldsymbol{v}=iD\left(\frac{\nabla\psi^{*}}{\psi^{*}}-\frac{\nabla\psi}{\psi}\right),\;\boldsymbol{u}=D\left(\frac{\nabla\psi^{*}}{\psi^{*}}+\frac{\nabla\psi}{\psi}\right).\label{3-20}
\end{equation}
The Schrödinger equation proper is obtained from (\ref{3-16}) by
taking
\begin{equation}
D=\hbar/2m.\label{3-22}
\end{equation}

\subsection{A possible connection with hydrodynamics}

By subtracting Eq. (\ref{3-14}) from (\ref{3-12}) and introducing
the \emph{access} (or backward) velocity, which is the linear combination
of the velocities $\boldsymbol{v}$ and $\boldsymbol{u}$ 
\begin{equation}
\boldsymbol{v}_{a}=\boldsymbol{v}-\boldsymbol{u}=\left(\mathscr{\mathcal{D}}_{c}-\mathcal{D}_{s}\right)\boldsymbol{x},\label{3-24}
\end{equation}
one obtains after some rearrangements 
\begin{equation}
\frac{\partial}{\partial t}\boldsymbol{v}_{a}+\boldsymbol{v}_{a}\cdot\nabla\boldsymbol{v}_{a}-D\nabla^{2}\boldsymbol{v}_{a}=-\frac{1}{m}\nabla(V+2V_{Q}),\label{3-26}
\end{equation}
where $V_{Q}$ stands for the quantum potential, 
\begin{equation}
V_{Q}=-\frac{1}{2}m\boldsymbol{u}^{2}-mD\boldsymbol{\nabla\cdot u}=-2mD^{2}\frac{\nabla^{2}\sqrt{\rho}}{\sqrt{\rho}}.\label{3-28}
\end{equation}
Incidentally, this equation shows that the quantum potential $V_{Q}$
is determined by the spatial density $\rho(x,t)$, a function that
depends on what is happening in the entire space, and thus contains
nonlocal information. Equation (\ref{3-26}) corresponds in hydrodynamics
to the Navier-Stokes equation for an incompressible, viscous fluid,
if $mD$ is taken for the kinematic viscosity $\nu$ and $(V+2V_{Q})/m$
is identified with the total pressure divided by the fluid density
$\rho_{o}$. With this identification, Planck's constant corresponds
to $\hbar\Longleftrightarrow2\rho_{o}\nu.$ Interestingly, however,
this is not an equation for the flow velocity $\boldsymbol{v}$ but
for the access velocity $\boldsymbol{v}_{a}$. This velocity represents
the coarse time-scale local average of the displacement from time
$t-\varDelta t$ to time $t$, i. e.,

\begin{equation}
\boldsymbol{v}_{a}(\boldsymbol{x})=\frac{\left\langle \boldsymbol{x}(t)-\boldsymbol{x}(t-\varDelta t)\right\rangle }{\Delta t}=\frac{\left\langle \Delta_{-}\boldsymbol{x}\right\rangle }{\Delta t},\label{3-30}
\end{equation}
where the average is taken over the ensemble of particles that cross
the point $\boldsymbol{x}$ at time $t$, taking into account that
at an earlier time $t-\varDelta t$ those particles had a distribution
of positions $\boldsymbol{x}^{\prime}=\boldsymbol{x}(t-\varDelta t)$.
It should be borne in mind that $\varDelta t$ must be much smaller
than the characteristic time of the systematic motion, but large enough
as to embrace the most closely spaced, rapid changes in $\boldsymbol{x}$.
If the system satisfies an ergodic principle, $t$ may also represent
the different times at which the same particle crosses the point $\boldsymbol{x}$
again and again.

A question that comes to mind is whether th\textsc{is }analogy between
the quantum and the hydrodynamic cases can be extended to include
the effect of a walking droplet on the fluid and, ultimately, obtain
the 'quantumlike' behaviour of the fluid-droplet system \cite{Bush15}.
In particular, are the statistics of the horizontal trajectories of
walking droplets the equivalent of the quantum statistics represented
by $\rho$, as is suggested in several papers on the subject?

For a system composed of a fluid layer acted on by a vibrating force
and by the bouncing droplet, the total pressure appearing in the Navier-Stokes
equation must include the external pressure $P$ on the fluid surface
due to the bouncing droplet, and the gravity term $g$ must include
the effective acceleration due to the vibrations, $g(t)=g+\gamma(t)$.
Further, the resulting Navier-Stokes equation has to be complemented
with the equation of motion for the droplet under the action of the
fluid. Since with each bouncing, the droplet pressure on the fluid
surface modifies the value of $h(\boldsymbol{x})$, and the dynamics
of the droplet depends in its turn on the force exerted upon it by
the modified fluid surface, one ends up with a coupled, nonlinear
system of equations that is difficult to analyze with a view to establishing
a (direct) comparison with \textsc{sqm}, viz. \textsc{qm}.

The accumulated memory effects on the wave field, and the dependence
of the droplet-fluid interaction on the relative phase and varying
shape of the fluid surface during contact, seem to be essential points
to consider in this regard. The first of these suggests focusing on
hydrodynamic analogs of the stationary states in quantum mechanics,
when the wave field has become stabilized and a well-defined histogram
of the droplet positions is obtained. The second point suggests introducing
a random element in the description of the horizontal motions, which
could serve to bring to the surface the counterpart of the diffusion
coefficient of \textsc{sqm} -- or equivalently, of the stochastic
velocity $\boldsymbol{u}$. It seems to us of considerable importance
that $\boldsymbol{u}$ may acquire values comparable to those of $\boldsymbol{v.}$

To give precision to the above considerations, we rewrite Eq. (\ref{3-26})
in the form 
\begin{equation}
\frac{D\boldsymbol{v}_{a}}{Dt}=-\frac{1}{m}\nabla(V-mD\boldsymbol{\nabla\cdot v}_{a}+2V_{Q}).\label{3-32}
\end{equation}
This equation, which is just another form of the Schrödinger equation,
offers a statistical description of the moving quantum \emph{particles}
of the problem under scrutiny, the analog of the droplets. We should
then perhaps reconsider our previous 'hydrodynamic' point of view
and take equation (\ref{3-32}), and thus (\ref{3-26}), as the quantum
equivalent (or analog) of Newton's equation of motion for the walking
droplets. The mean local force acting on the droplets contains then,
in addition to the expected classical components, a term similar to
the quantum potential with its associated nonlocality. 

\section{The wave element in SED}

A usual starting point for the analysis of the particle dynamics in
\noun{sed} is the (nonrelativistic) equation of motion, known as Braffort-Marshall
equation 
\begin{equation}
m\ddot{\boldsymbol{x}}=\boldsymbol{f}(\boldsymbol{x})+m\tau\boldsymbol{\dddot{x}}+e\boldsymbol{E}(t),\label{4-2}
\end{equation}
where $m\tau\boldsymbol{\dddot{x}}$ stands for the radiation reaction
force, with $\tau=2e^{2}/3mc^{3}$ \cite{Dice}. For an electron,
$\tau\approx10^{-23}$ s. $\boldsymbol{E}(t)$ represents the electric
component of the \noun{zpf} taken in the long-wavelength approximation,
with time correlation given in the continuum limit by ($j,k=1,2,3$)
\begin{subequations} \label{EE}
\begin{equation}
\left\langle E_{k}(s)E_{j}(t)\right\rangle =\delta_{kj}\varphi(t-s),\label{4-4}
\end{equation}
 where the spectral function
\begin{equation}
\varphi(t-s)=\frac{2\hbar}{3\pi c^{3}}\intop_{0}^{\infty}d\omega\,\omega^{3}\cos\omega(t-s)\label{4-6}
\end{equation}
\end{subequations} corresponds to an energy $\hbar\omega/2$ per
mode. This establishes a crucial distinction between classical, Brownian-type
processes, governed by a white noise, and the present stochastic process
which is governed by a stationary, correlated (wave) field with a
highly coloured spectrum. 

A statistical treatment of the one-particle problem, starting from
Eq. (\ref{4-2}) and involving the construction of a generalized Fokker-Planck
equation, has been shown to lead in the radiationless approximation
to the Schrödinger equation \cite{PeCeVaEta14,TEQ}; this is the approximation
that corresponds to quantum mechanics. The stationary states are obtained
as the solutions that satisfy the energy-balance condition, as predicted
by Nernst over a century ago. When the radiative terms are not neglected,
the theory reproduces in addition the corresponding corrections in
coincidence with non-relativistic quantum electrodynamics \cite{NosRMF13,TEQ}.
In particular, the formulas for the radiative lifetimes are obtained
for the excited states that in the Schrödinger approximation appear
as stationary. 

Interestingly, in the process leading to the Schrödinger equation
via the Fokker-Planck equation, which focuses on the statistical properties
of the dynamics, the wave element is not conspicuous; it remains as
concealed as in \noun{sqm}. Yet the \noun{zpf} eventually leaves its
indelible mark through the appearance of Planck's constant in the
Schrödinger 'wave' equation. 

An alternative route followed in a subsequent development of \noun{sed},
called linear stochastic electrodynamics (\noun{lsed}), takes us to
the quantum equations in its matrix formulation (\cite{TEQ}; see
also \cite{QSMF2021,FOOP2022} for more recent work). As is well known,
the wave element is absent from the final (Heisenberg) equations,
which have a purely mechanical aspect. However, in the \noun{sed}
process leading to these equations (see section \ref{CtoQ} below),
the relevant field modes with which the particle interacts appear
explicitly. This allows us to investigate the role played by such
field modes when two (not directly interacting) identical particles
are simultaneously connected to them. As a result, the entanglement
of particles finds an explanation in the correlation between their
motions established through the common field modes \cite{NosEntangl,TEQ}.
In this regard, it is appropriate to mention the connection with the
hydrodynamic work by Borghesi et al \cite{Borgh14}, who investigate
experimentally the energy stored in the wave field for two coupled
walking droplets and how it conveys an interaction between them. More
generally, the observation that the background field acts as a bridge
between particles has important consequences for the statistics of
(identical) quantum particles. 

\subsection{Compton's frequency revindicated}

In de Broglie's pioneering work on wave mechanics, the Compton frequency
is well known to have played a key role as the frequency of the particle's
internal clock. (Later it was understood by some that de Broglie's
wavelength has actually a statistical meaning.) Nevertheless, the
nature of the associated wave in de Broglie's theory, which was basic
for the development of Schrödinger's theory, was left unidentified
by de Broglie, and in fact it remains to date concealed at the core
of the foundations of quantum mechanics.

A variety of recent works, both theoretical and experimental, point
to a revival of the internal clock conjectured by de Broglie, which
in turn may have an important impact on our understanding of the quantum
phenomenon. Among such works we have on one hand a series of experiments
in which high-energy electron beams are channelled through silicon
crystals \cite{Cat2008,Osche2011}, providing evidence of a resonance
suggestive of de Broglie's clock. Ironically, these findings have
received apparently almost no attention, except for an essay by Hestenes
\cite{Hest2008} and a Monte-Carlo-based analysis by Bauer \cite{Bauer}
showing consistency with Dirac's description of the free particle.

On the other hand there is a growing series of both experimental and
theoretical work carried out in the field of bouncing droplets referred
to above \cite{Bush15}, in which a high-frequency vibration ---somehow
parallel to de Broglie's clock--- is shown to play a central role
in producing phenomena that suggest a hydrodynamic quantum analogy.
Also in other fields of physics, high-frequency vibrations imposed
on a material system seem to have an important influence, inducing
qualitative changes in the dynamics of the system. Worth mentioning
are a couple of recent extensions to a generic elastic system with
a bead \cite{Borgh}, where also the internal clock of high frequency
appears as a key ingredient.

De Broglie's frequency has had occasional appearances in \textsc{sed}.
A first attempt to establish contact with it is the work by Surdin
\cite{Surdin82} where the electromagnetic nature of de Broglie's
wave is specified as follows from \textsc{sed}, and some experiments
dealing with interference and diffraction of neutrons are considered.
In \cite{Dice,TEQ} we present a preliminary view on the functioning
of the modes of the \textsc{zpf} of Compton's frequency, and their
relation to de Broglie's clock, with a few simple but interesting
results. In dealing with some quantum problems within the \textsc{sed}
framework we have been induced to introduce, as an ancillary step,
a revised version of de Broglie's idea, appropriately updated in accordance
with the demands of \textsc{sed}, which means assigning an \emph{electromagnetic}
nature to the de Broglie wave. In the process leading to the appearance
of de Brogle's wave, the assumption of a resonant interaction of the
electron with the \textsc{zpf} modes of Compton's frequency has been
the starting point. Essentially this amounts to distinguishing the
modes of the \textsc{zpf} of Compton's frequency from the rest of
the field. 

More recently, we have put forward an estimate of the duration of
a transition between atomic states---the infamous quantum jumps---based
on the assumption that the transition is triggered precisely by a
resonance of the atomic electron with modes of the zero-point radiation
field of Compton's frequency \cite{Nos PLA20}. The theoretical result,
given essentially by the expression $(\alpha\omega_{C})^{-1}$, where
$\alpha\sim1/137$ is the fine structure constant and $\omega_{C}$
the Compton angular frequency for the electron, lies well within the
range of the experimentally estimated values, which is of the order
of attoseconds (10$^{-18}$ s) \cite{Oss2,Minev}. Incidentally, one
can still come across articles negating quantum transitions, in adherence
to Bohr's dictum---and any other kind of apparent discontinuities,
for that matter (e.g. \cite{Jad14})---or taking them as a sudden
increase of our knowledge of the system (e. g., \cite{Zeh,Plenio})
rather than a physical phenomenon. 

\section{On the classical-to-quantum transition\label{CtoQ}}

The transition from the classical to the quantum regime represents
a most delicate point of the emerging-quantum theory. A detailed account
of the dynamics leading to quantization, and of the conditions under
which quantization is obtained, is still a pending task, although
some important steps have been achieved. 

Notably, by applying a Hamiltonian treatment to the composite (initially
noninteracting) particle-field system, it is possible to show that
in the classical-to-quantum transition, the field takes control of
the response of the particle. As a result of the interplay of stochasticity
and dissipation, the particle loses memory of its initial conditions
and ends up responding linearly and resonantly to a well-defined set
of field modes, which depends in each instance on the specific problem
(i. e., on the external forces acting on the particle) \cite{QSMF2021,FOOP2022}.
In the new situation the dynamics is no longer described by the phase-space
mechanical variables, but by the coefficients of the response of the
particle to those field modes. The response coefficients turn out
to be nothing less than the matrix elements of the operators $\hat{x}$
and $\hat{p}$, which satisfy the basic quantum commutator $\left[\hat{x},\hat{p}\right]=i\hbar$,
just as was advanced by Heisenberg in his pioneering work on quantum
mechanics \cite{Heis25}; this reflects the fact that the Hamiltonian
evolution preserves the original symplectic structure. 

The Hilbert-space formalism represents thus a compact and elegant
way of describing the response of the particle to a set of (relevant)
field modes once the quantum regime has been attained, at the cost
of a space-time description of what is really happening inside the
atom. Interestingly, the field has disappeared once more from the
picture, this time leaving its indelible mark through the appearance
of Planck's constant in the basic commutator. 

It should be noted that prior to the onset of the quantum regime,
the dynamics is irreversible and memory accumulates in the near background
field, which is an indication of a highly non-Markovian process. In
the quantum regime, by contrast, the stochastic process is Markovian,
the Markovianity referring to the loss of memory between quantum-mechanical
processes such as atomic transitions. This feature is related with
the times involved in the description: we recall from section \ref{SQM}
that the time interval $\Delta t$ introduced in \noun{sqm} to derive
the \noun{qm} equations entails a statistical coarse-grain description
of the dynamics. The minimum times involved in (non-relativistic)
\noun{qm} are smaller than the orbital periods and the lifetimes of
excited states, but certainly larger than the inverse of the Compton
(or Zitterbewegung) frequency, or even the time involved in a 'quantum
jump' (which as discussed above, is an estimated 100 times larger
than the Compton time). For this reason, \noun{qm} as we know it is
unable to describe with a sufficiently high resolution in time the
trajectories of particles, transitions between states, or other continuous
processes. 

\section{Final comments}

The experience with \noun{sed}, as illustrated by the results discussed
in the preceding sections, is that the main features characteristic
of quantum-mechanical systems emerge from the permanent interaction
of the random zero-point radiation field with matter; for this reason
it can be said that \noun{qm} is an emergent theory. By adding the
\noun{zpf} to the quantum ontology, a physically coherent, local and
objective picture of the quantum phenomenon is obtained, free of ad-hoc
postulates and mysterious elements.

It seems hard to rule out the possibility that analogous quantization
phenomena take place in other realms of physics, in which a stationary,
vibrating background acts permanently on one or more corpuscles. The
(hydrodynamic) walking-droplet system and the (electrodynamic) atomic-matter
system are perhaps two prominent examples of a more general phenomenon
with interesting ramifications. It remains to be seen whether a common
ground can be found that allows us to deepen our mutual learning,
further identify commonalities, and establish the extent of the analogies,
for the benefit of a more comprehensive, satisfactory understanding
of the quantum phenomenon.

\end{document}